\documentstyle[12pt]{article}
\begin{document}

\title{Mean Field Methods for Atomic and Nuclear Reactions:\\
The Link between Time--Dependent and Time--Independent Approaches}
\author{J. Uhlig, J. C. Lemm and A. Weiguny\\
Institut f\"ur Theoretische Physik I\\
Universit\"at M\"unster\\
D--48149 M\"unster, Germany}
\date{}
\maketitle

$\,\,\,$ PACS: 03.65.Nk; 24.10.-i; 34.10.+x

\begin{abstract}
Three variants of mean field methods for atomic and nuclear reactions
are compared with respect to both conception and applicability: The
time--dependent Hartree--Fock method solves the equation of motion for a
Hermitian density operator as initial value problem, with the colliding
fragments in a continuum state of relative motion. With no specification
of the final state, the method is restricted to inclusive reactions. The
time--dependent mean field method, as developed by Kerman, Levit and
Negele as well as by Reinhardt, calculates the density for specific
transitions and thus applies to exclusive reactions. It uses the
Hubbard--Stratonovich transformation to express the full time--development
operator with two--body interactions as functional integral over
one--body densities. In stationary phase approximation and with Slater
determinants as initial and final states, it defines non--Hermitian,
time--dependent mean field equations to be solved self--consistently as
boundary value problem in time. The time--independent mean field method
of Giraud and Nagarajan is based on a Schwinger--type variational
principle for the resolvent. It leads to a set of inhomogeneous,
non--Hermitian
equations of Hartree--Fock type to be solved for given total energy. All
information about initial and final channels is contained in the
inhomogeneities, hence the method is designed for exclusive reactions. A
direct link is established between the time--dependent and
time--independent versions. Their relation is non--trivial due to the
non--linear nature of mean field methods.
\end{abstract}
\thispagestyle{empty}
\pagebreak
\setcounter{page}{1}
\section{Similarities and Differences}
After the success which mean field methods have had for bound state
problems in various fields of physics, it was only natural to try the
mean field concept for scattering states as well. The original attempt
in this direction is the time--dependent Hartree--Fock method (TDHF)
developed about 20 years ago [1]. In this method one solves the
equation of motion for the one--body density operator $\rho\,=\,\rho\,(t)$,
\renewcommand{\theequation}{1.\arabic{equation}}
\begin{equation}
i\,\hbar\,\frac{\partial}{\partial\,t}\,\rho\,=\,[h,\,\rho]\,\,,
\end{equation}
with an \underline{initial} condition for $\rho$,
\begin{equation}
\rho\,(t_{i})\,=\,\rho_{i}\,\,,
\end{equation}
describing an unbound state of relative motion of the colliding
fragments. The density operator
$\rho$ can
be represented in just \underline{one} basis of single--particle states
$\psi_{m}\,(t)$,
\begin{equation}
\rho\,=\,\sum\limits_{m\,=\,1}^{N}\mid\psi_{m}\,(t)\,><\,\psi_{m}\,(t)\mid\,\,
\end{equation}
for $N$ particles and with $\psi_{m}\,(t)$ orthonormalized. The
single--particle Hamiltonian $h$ is Hermitian and has standard 
Hartree--Fock structure, $h\,=\,t\,+\,u$ with mean field potential
\begin{equation}
u\,=\,\sum\limits_{m\,=\,1}^{N}\,<\cdot\,\psi_{m}\,(t)\,|\,v\,|\,
\cdot\,\psi_{m}\,(t)>\,\,,
\end{equation}
where the dots represent wave functions of some arbitrary
single--particle basis. The single--particle states $\psi_{m}\,(t)$ are
determined from the time--dependent Hartree--Fock equations
\begin{equation}
i\,\hbar\,\frac{\partial}{\partial\,t}\,|
\,\psi_{m}\,(t)>\,\,=\,h\,|\,\psi_{m}\,(t)>
\end{equation}
with initial conditions
\begin{equation}
\psi_{m}\,(t_{i})\,=\,\chi_{m}\qquad{\rm for}\,\,\,
m\,=\,1,\,2,\,...\,N\,\,.
\end{equation}
Standard manipulation of eq. (1.5) and its Hermitian adjoint leads to
eq. (1.1), with initial condition (1.2) corresponding to (1.6). In
practice, the non--linear equations (1.1) and (1.5), respectively, are
solved by iteration: A set of $N$ occupied orbitals is used to compute
the potential $u$. With the corresponding solutions of (1.1) or (1.5)
one recalculates $u$ till self--consistency is reached. As each iteration
step requires as input only the set of $N$ functions $\psi_{m}$ at fixed
time $t$, the TDHF method as initial value problem is "local" in
time.\\[3ex]
From the densities $\rho_{b}(\vec{r},\,t)$
according to (1.1) and (1.2), taken at various classical impact
parameters $b$, one may calculate a classical cross section. Although it
can generate nice snapshots of the density distribution
during the scattering process, the method has two problems: First, as
initial value problem it can at best handle inclusive reactions, since
no specification of a final channel enters the formalism. The second,
more serious problem concerns ''spurious cross channel correlations'':
Starting from the determinant $\Psi\,(t_{i})$ corresponding to
$\rho_{i}$ the method generates a single determinant $\Psi\,(t)$.
Due to the non--linear nature of the TDHF equations, this
determinant $\Psi\,(t)$ continues to vary as time goes to infinity.
Hence, if this wave
function is expanded in an orthogonal set of channel wave functions, the
expansion coefficients will not be constant asymptotically. Thus an
S--matrix, constructed by projecting the TDHF wave functions onto
channel wave functions, would not be constant in time [2].\\[3ex]
The time--dependent mean field method (TDMF) [3, 4] also uses an
equation of motion like (1.1),
\begin{equation}
i\,\hbar\,\frac{\partial}{\partial\,t}\,\rho\,=\,[h,\,\rho]\,\,,
\end{equation}
however, there are two important differences:\\
1. The density operator
$\rho$ of (1.7) is expanded in \underline{two} sets of mutually
biorthogonal single--particle functions,
\begin{equation}
\rho\,=\,\sum\limits_{m\,=\,1}^{N}\,\frac{\mid\psi_{m}\,(t)\,><\,\tilde{\psi}_{m}\,
(t)\mid}{<\,\tilde{\psi}_{m}\,(t)\,|\,\psi_{m}\,(t)>}\,\,,
\end{equation}
and the Hamiltonian $h\,=\,t\,+\,u$ with mean field potential
\begin{equation}
u\,=\,\sum\limits_{m\,=\,1}^{N}\,\frac{<\cdot\,\tilde{\psi}_{m}\,(t)\,|\,v\,|\,\cdot\,\psi_{m}\,(t)>}{<\tilde{\psi}_{m}\,|\,\psi_{m}>}
\end{equation}
is \underline{non}--Hermitian in general.\\
2. Eq.(1.7) has to be solved self--consistently as \underline{boundary}
problem in time $t$, fixing
\begin{equation}
\rho\,(t_{i})\,=\,\rho_{i}\quad{\rm and}\quad
\rho\,(t_{f})\,=\,\rho_{f}\,\,.
\end{equation}
The single--particle functions $\psi_{m}\,(t)$ are obtained from
\begin{equation}
i\,\hbar\,\frac{\partial}{\partial\,t}\,|\,\psi_{m}\,(t)>\,\,=\,
h\,|\,\psi_{m}\,(t)>
\end{equation}
by forward propagation of initial wave functions
\begin{equation}
\psi_{m}\,(t_{i})\,=\,\chi_{m}\qquad {\rm for}\,\,\,
m\,=\,1,\,2,\,...\,N\,\,.
\end{equation}
Analogously $\tilde{\psi}_{m}\,(t)$ results from
\renewcommand{\theequation}{1.11'}
\begin{equation}
-\,i\,\hbar\,\frac{\partial}{\partial\,t}\,<\tilde{\psi}_{m}\,(t)\,|\,
=\,<\tilde{\psi}_{m}\,(t)\,|\,h
\end{equation}
by backward propagation of
\renewcommand{\theequation}{1.12'}
\begin{equation}
\tilde{\psi}_{m}\,(t_{f})\,=\,\chi^{'}_{m}\,\,
\qquad {\rm for}\,\,\,
m\,=\,1,\,2,\,...\,N\,\,.
\end{equation}
Combining (1.11) and (1.11') in the usual way, one obtains (1.7) with
conditions (1.10) from (1.12), (1.12'). One also proves easily
that
\renewcommand{\theequation}{1.13}
\begin{equation}
\frac{\partial}{\partial\,t}\,<\tilde{\psi}_{m}|\,\psi_{n}>\,\,=\,0\,\,,
\end{equation}
hence if $\tilde{\psi}_{m},\,\psi_{n}$ are chosen biorthogonal
at $t\,=\,t_{i}$, they will remain so at any time. In general,
$\tilde{\psi}_{m}$ and $\psi_{m}$ will be not complex--conjugate
to each other, and $\rho$ will be non--Hermitian. Only when
$\chi_{m}^{'}$ is generated from $\chi_{m}$ by the mean field time
development operator $U_{h}\,(t_{f}\,-\,t_{i})$, we have
$\tilde{\psi}_{m}\,(t)\,=\,\psi_{m}^{*}\,(t)$ in which case
TDMF reduces to TDHF with just one set of single--particle functions.
Equations (1.7), (1.10) as well as the coupled equations (1.11), (1.11')
together with (1.12), (1.12') constitute a boundary condition problem in
which the time plays a role similar to that of the spatial
coordinates [5].\\[3ex]
In practice, one may try to solve this problem self--consistently in
analogy to the static Hartree--Fock problem by iteration [5] which
involves an initial guess of the single--particle functions
$\tilde{\psi}_{m}\,(t),\,\psi_{m}\,(t)$ for \underline{all}
times $t$ between $t_{i}$ and $t_{f}$. In this sense, the TDMF boundary
condition problem is highly "non--local" in the variable $t$.
Since initial \underline{and} final states are taken care of, the method is
able to describe also exclusive reactions: Each quantum process, leading
from given initial state $\chi$ to some final state $\chi^{'}$, has its
own time--dependent mean field assigned. It has been proven [2]
that, within the framework of TDMF, an S--matrix can be defined which
becomes asymptotically constant. The problem with the TDMF approach lies
in the above non--locality in time, for which there seems to exist no
practicable algorithm for
actual numerical calculations of (3 + 1)--dimensional systems.\\[3ex]
Some comments on the lack of Hermiticity of $\rho$ and of $h\,(\rho)$
are in order. This non--Hermitian structure is of a special type which
preserves the usual properties of a single--particle density matrix
connected to a Slater determinant. From the explicit expression (1.8) we
have immediately
\renewcommand{\theequation}{1.14}
\begin{equation}
{\rm Tr}\,\rho\,=\,N\qquad {\rm and} \qquad
\rho^{2}\,=\,\rho\,\,,
\end{equation}
hence particle number is fixed and $\rho$ is a projector on the space of
occupied orbitals. Moreover, $|\,\psi_{m}>$ and
$<\tilde{\psi}_{m}|$ are right and left eigenstates of $\rho$,
\renewcommand{\theequation}{1.15}
\begin{equation}
\rho\,|\,\psi_{m}\,(t)>\,\,=\,N_{m}\,|\,\psi_{m}\,(t)>\,\,;\,\,
<\tilde{\psi}_{m}\,(t)\,|\,\rho\,=\,<\tilde{\psi}_{m}\,(t)\,|\,N_{m}
\end{equation}
with eigenvalues
\renewcommand{\theequation}{1.16}
\begin{equation}
N_{m}\,=\,\left\{
\begin{array}{ll}
1\,\,\,{\rm for}\,\,m\,\,{\rm occupied}\\
0\,\,\,{\rm otherwise}
\end{array}\right.\,\,.
\end{equation}
The eigenvalues of $h\,(\rho)$ will be complex in general, hence
$\psi_{m}$ and $\tilde{\psi}_{m}$ correspond to quasi--particles
of finite lifetime. They represent intermediate states which the system
passes during the reaction process. The S--matrix of the mean field
approach will always violate unitarity, irrespective of the fact whether
$h\,(\rho)$ is Hermitian or not: The ${\cal S}$--operator depends on $\rho$ which
in turn depends on the initial and final
states of a specific
reaction as mentioned above. Each quantum process $\chi\,\to\,\chi^{'}$
has its own mean field, calculated without reference to any other
possible process.\\[3ex]
An alternative to TDMF is the time--independent mean field method (TIMF)
based on a Schwinger--type variational principle [6, 7] as described below in section 2. Like TDMF it uses
\underline{two} sets of variational functions and leads to a set of
inhomogeneous equations for the single--particle functions
$\varphi_{i}^{'},\,\varphi_{i}$ which may be chosen biorthogonal. The
corresponding density operator
\begin{equation}
\renewcommand{\theequation}{1.17}
\rho\,=\,\sum\limits_{i\,=\,1}^{N}\frac{\mid\varphi_{i}><\varphi_{i}^{'}\mid}{<\varphi_{i}^{'}\mid\varphi_{i}>}
\end{equation}
obeys an inhomogeneous equation,
\begin{equation}
\renewcommand{\theequation}{1.18}
[h,\,\rho]\,=\,\overline{\eta}\,(\rho_{\chi^{'}}\,-\,\rho_{\chi})\quad;\quad
\overline{\eta}\,=\,E\,-\,\frac{<\phi^{'}\mid H\mid\phi >}
{<\phi^{'}\mid\phi>}\,\,,
\end{equation}
where $E$ is the total energy of the system, $\phi,\,\phi^{'}$ are Slater
determinants built from the single--particle functions
$\varphi_{i},\,\varphi_{i}^{'}$ 
and $\rho_{\chi^{'}},\,\rho_{\chi}$ 
are ''mixed'' densities containing 
all information about initial and final states $\chi$, $\chi^{'}$.
Hence the method can describe exclusive reactions. As in TDMF
$h$ is
\underline{non}--Hermitian,
\begin{equation}
\renewcommand{\theequation}{1.19}
h\,=\,\sum\limits_{i\,=\,1}^{N}\,\frac{<\cdot\,\varphi_{i}^{'}\,|\,v\,|\,
\cdot\,\varphi_{i}>}{<\varphi_{i}^{'}\,|\,\varphi_{i}>}\,,
\end{equation}
but now of course time--independent as are
the single--particle functions $\varphi_{i},\,\varphi^{'}_{i}$. Their self--energies $\eta_i$ are complex in general due to the lack of Hermiticity of $h$. As
regards its practical applicability [8], TIMF is much simpler than TDMF: One
has to solve inhomogeneous, complex equations of Hartree--Fock type at
some given energy $E$. In this respect TIMF is comparable to
TDHF: While TDHF is ''local'' in time, TIMF is ''local'' in energy.\\[3ex]
In view of the above similarities and differences of time--dependent and
time--independent mean field methods (see table 1), it appears desirable to establish a direct link between TDMF and TIMF. Their relation is non--trivial due to
the non--linear nature of mean field methods.\\[3ex]
The paper is organized as follows: In section 2 we shall derive matrix
elements of the resolvent operator as stationary values of a
Schwinger--type functional in mean field approximation. Its stationarity
conditions are given by the TIMF equations (1.17) and (1.18). To
calculate matrix elements of the exact time--development operator we use
a path integral representation based on the Hubbard--Stratonovich
transformation (section 3). Ambiguities of the respective auxiliary
fields $\sigma$ are briefly discussed in section 4, following [9]. The
mean field approximation of the time--development operator is then
obtained by applying the stationary phase approximation to the path
integrals over the auxiliary fields $\sigma$  in section 5. The result is the
boundary condition problem for $\rho$, eqs. (1.7) to (1.10), of the TDMF
method. The link between the two methods is obtained in two steps:
First, in static approximation of TDMF (section 6) we can introduce
single--particle propagators $g_{m}\,=\,(w_{m}\,-\,h(\sigma))^{-1}$ by Fourier
transformation of the corresponding time--development operator
$\exp\,\{-\,\frac{i}{\hbar}\,h(\sigma)\,T\}$. Second, the Fourier integrals over $w_{m}$ are solved in stationary phase approximation on the same footing
(section 7) as the path integrals over the static auxiliary fields $\sigma$. Thus the integrals over real variables  $w_{m}$ are replaced by the integrand
at some complex values $\omega_m^{\circ}$ which are identified in section 9 as the
self--energies $\eta_{m}$ of TIMF after the $T$--integration has been carried
out (section 8). At the same time one reproduces the TIMF value of the
resolvent. In conclusion (section 9): TIMF turns out to be a static
approximation of TDMF in semi--classical approximation.

\section{Resolvent operator in mean field approximation}
The TIMF method solves the stationarity equations of a functional like [6, 7]
\setcounter{equation}{0}
\renewcommand{\theequation}{2.\arabic{equation}}
\begin{equation}
F\,(\Psi^{'},\,\Psi)\,=\,<\chi^{'}|\Psi>\,+\,<\Psi^{'}|\chi>\,-\,<\Psi^{'}
|(E\,-\,H)|\Psi>
\end{equation}
which read
\begin{equation}
(E\,-\,H)|\Psi>\,=\,|\chi>\,\,;\,\,<\Psi^{'}|(E\,-\,H)\,=\,<\chi^{'}|
\end{equation}
with $\chi,\,\chi^{'}$ as initial and final states. The resolvent matrix
element between $\chi$ and $\chi^{'}$ is then given by
\begin{equation}
<\chi^{'}|(E\,-\,H)^{-1}|\chi>\,=\,<\chi^{'}|\Psi>\,=\,<\Psi^{'}|\chi>\,\,.
\end{equation}
In mean field approximation one takes $\chi,\,\chi^{'}$ as Slater
determinants,
\begin{equation}
\chi\,=\,{\cal{A}}\,\prod\limits_{i\,=\,1}^{N}\,\chi_{i}\,\,\,\,;\,\,\,\,
\chi^{'}\,=\,{\cal{A}}\,\prod\limits_{i\,=\,1}^{N}\,\chi_{i}^{'}\,\,,
\end{equation}
and the variational functions $\Psi,\,\Psi^{'}$ are correspondingly
restricted to Slater determinants as well,
\begin{equation}
\phi\,=\,{\cal{A}}\,\prod\limits_{i\,=\,1}^{N}\,\varphi_{i}\,\,\,\,;\,\,\,\,
\phi^{'}\,=\,{\cal{A}}\,\prod\limits_{i\,=\,1}^{N}\,\varphi^{'}_{i}\,\,.
\end{equation}
The antisymmetrizer ${\cal{A}}$ is defined as
${\cal{A}}\,=\,(N\,!)^{-\frac{1}{2}}\,\sum\limits_{P}\,(-)^{P}\,P$. One
then has to vary
\begin{equation}
F\,(\phi^{'},\phi)\,=\,<\chi^{'}|\phi>\,+\,<\phi^{'}|\chi>\,-\,
<\phi^{'}|(E\,-\,H)|\phi>\,\,.
\end{equation}
To calculate $F\,(\phi^{'},\phi)$ and its functional derivatives, one
can use the invariance of the above four determinants under unitary
transformation of their orbitals in order to diagonalize the four
$N$--dimensional matrices
$<\chi_{i}^{'}\mid\varphi_{j}>,\,<\varphi_{i}^{'}\mid\chi_{j}>,\,
<\varphi_{i}^{'}\mid\varphi_{j}>$ and $<\varphi_{i}^{'}\mid h\mid\varphi_{j}>$.
Here $h=t+u$ is the single--particle Hamiltonian with $u$ the
Hartree--Fock potential in the biorthogonal basis
$\varphi_{i}^{'},\varphi_{j}$. In this representation the stationarity
equations of (2.6) read:
\begin{equation}
(\eta_{i}\,-\,h)\mid\varphi_{i}>\,=\,\lambda_{i}\mid\chi_{i}>\,\,\,;\,\,\,
<\varphi_{i}^{'}\mid (\eta_{i}\,-\,h)\,=\,<\chi^{'}_{i}\mid\lambda_{i}^{'}
\end{equation}
with complex self--energies
\begin{equation}
\eta_{i}\,=\,E\,-\,\frac{<\phi^{'}\mid H\mid\phi>}{<\phi^{'}\mid\phi>}\,+\,
\frac{<\varphi_{i}^{'}\mid h\mid\varphi_{i}>}{<\varphi_{i}^{'}\mid\varphi_{i}>}\,\,.
\end{equation}
The stationary value of $F\,(\phi^{'},\,\phi)$ is then
\begin{equation}
F_{stat}\,=\,<\chi^{'}\mid\phi>\,=\,<\phi^{'}\mid\chi>\,=\,\prod\limits_{i}\,
\lambda_{i}<\chi^{'}_{i}\mid g_{i}\mid\chi_{i}>\,=\,\prod\limits_{i}\,\lambda_{i}^{'}
<\chi_{i}^{'}\mid g_{i}\mid\chi_{i}>
\end{equation}
with single--particle propagators
\begin{equation}
g_{i}\,=\,(\eta_{i}\,-\,h)^{-1}\,\,.
\end{equation}
The explicit form of $\lambda_{i},\,\lambda_{i}^{'}$ --  which we shall
need in the following -- is
\begin{equation}
\lambda_{i}\,=\,\frac{<\phi^{'}\mid\chi>}{<\phi^{'}\mid\phi>}\,\cdot\,
\frac{<\varphi^{'}_{i}\mid\varphi_{i}>}{<\varphi_{i}^{'}\mid\chi_{i}>}\,\,\,;\,\,\,
\lambda_{i}^{'}\,=\,\frac{<\chi^{'}\mid\phi>}{<\phi^{'}\mid\phi>}\,\cdot\,
\frac{<\varphi_{i}^{'}\mid\varphi_{i}>}{<\chi^{'}_{i}\mid\varphi_{i}>}\,\,.
\end{equation}
They are related to each other and to $\overline{\eta}$ of eq. (1.18):
From (2.11) and (2.9) one finds
\begin{equation}
\prod\limits_{i}\,\lambda_{i}\,=\,\left(\frac{<\phi^{'}\mid\chi>}{<\phi^{'}\mid\phi>}
\right)^{N\,-\,1}\,=\,\left(
\frac{<\chi^{'}\mid\phi>}{<\phi^{'}\mid\phi>}\right)^{N\,-\,1}\,=\,
\prod\limits_{i}\,\lambda_{i}^{'}\,\,.
\end{equation}
Furthermore
\begin{equation}
\overline{\eta}^{N\,-\,1}\,=\,
\prod\limits_{i}\,\lambda_{i}\,=\,\prod\limits_{i}\,\lambda_{i}^{'}\,\,,
\end{equation}
as is easily seen: Scalar multiplication of (2.7) by $<\varphi_{i}^{'}\mid$
and
$\mid\varphi_{i}>$, respectively, gives together with (1.8) and (2.8):
\begin{equation}
\overline{\eta}\,=\,E\,-\,\frac{<\phi^{'}\mid H\mid\phi>}{<\phi^{'}\mid\phi>}\,=\,\eta_{i}\,-\,
\frac{<\varphi_{i}^{'}\mid h\mid\varphi_{i}>}{<\varphi_{i}^{'}\mid\varphi_{i}>}\,=\,\lambda_{i}\,
\frac{<\varphi_{i}^{'}\mid\chi_{i}>}{<\varphi_{i}^{'}\mid\varphi_{i}>}\,=\,\lambda_{i}^{'}\,
\frac{<\chi^{'}_{i}\mid\varphi_{i}>}{<\varphi_{i}^{'}\mid\varphi_{i}>}
\end{equation}
and with (2.11)
\begin{equation}
\overline{\eta}\,=\,
\frac{<\phi^{'}\mid\chi>}{<\phi^{'}\mid\phi>}\,=\,
\frac{<\chi^{'}\mid\phi>}{<\phi^{'}\mid\phi>}\,\,.
\end{equation}
Combining (2.12) and (2.15) then results in (2.13).\\[3ex]
It is now easy to obtain an equation for the density operator
\begin{equation}
\rho\,=\,\sum\limits_{i\,=\,1}^{N}\,
\frac{\mid\varphi_{i}><\varphi_{i}^{'}\mid}{<\varphi_{i}^{'}\mid\varphi_{i}>}\,\,,
\end{equation}
using biorthogonality but leaving the normalization of the
single--particle functions open. The dyadic
product of (2.7) with $<\varphi_{i}^{'}\mid$ and $\mid\varphi_{i}>$ from right
and left, respectively, gives
\begin{equation}
(\eta_{i}\,-\,h)\mid\varphi_{i}><\varphi_{i}^{'}\mid\,=\,\lambda_{i}\mid\chi_{i}><\varphi_{i}^{'}\mid\,\,\,\,;\,\,\,\,
\mid\varphi_{i}><\varphi_{i}^{'}\mid(\eta_{i}\,-\,h)\,=\,\mid\varphi_{i}><\chi_{i}^{'}\mid
\lambda_{i}^{'}\,\,.
\end{equation}
Dividing by the norm $<\varphi_{i}^{'}\mid\varphi_{i}>$, summing over $i$
and forming the difference finally results in:
\begin{equation}
[h,\,\rho]\,=\,\overline{\eta}\,(\rho_{{\chi}^{'}}\,-\,\rho_{\chi})\,\,,
\end{equation}
if one defines the "mixed" densities
\begin{equation}
\rho_{\chi^{'}}\,=\,\sum\limits_{i\,=\,1}^{N}\,
\frac{\mid\varphi_{i}><\chi_{i}^{'}\mid}{<\chi_{i}^{'}\mid\varphi_{i}>}\,\,\,\,;\,\,\,\,
\rho_{\chi}\,=\,\sum\limits_{i\,=\,1}^{N}\,
\frac{\mid\chi_{i}><\varphi_{i}^{'}\mid}{<\varphi_{i}^{'}\mid\chi_{i}>}
\end{equation}
and uses equation (2.14). This constitutes the proof of equation
(1.18).

\section{Path integral representation of the time--development
operator}
\setcounter{equation}{0}
\renewcommand{\theequation}{3.\arabic{equation}}
If a state $\chi$ is prepared at time $t_{i}$, then it will develop
with time upto $t\,=\,t_{f}$ as
\begin{equation}
|\chi\,(t_{f})>\,=\,U\,(t_{f}\,-\,t_{i})|\chi>
\end{equation}
with
\begin{equation}
U\,(t_{f}\,-\,t_{i})\,=\,\exp\,\left\{-\,\frac{i}{\hbar}\,(t_{f}\,-\,t_{i})\,H\right\}\,\,.
\end{equation}
The probability of finding some state $\mid\chi^{'}>$ at $t\,=\,t_{f}$ is
then obtained from the amplitude
\begin{equation}
<\chi^{'}|\chi(t_{f})>\,=\,<\chi^{'}|U\,(t_{f}\,-\,t_{i})|\chi>\,\,.
\end{equation}
An exact representation of the time--development operator of the
many--body
problem in terms of a time--dependent single--particle Hamiltonian can be
achieved
by the Hubbard--Stratonovich transformation. To this end one rewrites the
full Hamiltonian $H$ in terms of operators [3]
\begin{equation}
\rho_{\alpha\,\gamma}\,=\,a^{\dagger}_{\alpha}\,a_{\gamma}
\end{equation}
in arbitrary orthonormal basis as
\begin{equation}
H\,=\,\sum\limits_{\alpha,\,\gamma}\,K_{\alpha\,\gamma}\,\rho_{\alpha\,\gamma}\,+\,
\frac{1}{2}\,\sum\limits_{\alpha,\,\beta,\,\gamma,\,\delta}\,\rho_{\alpha\,\gamma}\,v_{\alpha\,\beta\,\gamma\,\delta}
\,\rho_{\beta\,\delta}
\end{equation}
with matrix elements
\begin{equation}
v_{\alpha\,\beta\,\gamma\,\delta}\,=\,\int\int\,d^{3}\,\vec{r}\,d^{3}\,\vec{r}^{\,'}\,
\varphi_{\alpha}^{\star}\,(\vec{r})\,\varphi_{\beta}^{\star}\,(\vec{r}^{\,'})\,v\,
(\vec{r},\,\vec{r}^{\,'})\,\varphi_{\gamma}\,(\vec{r})\,\varphi_{\delta}\,
(\vec{r}^{\,'})\,=\,(\alpha\,\beta|v|\gamma\,\delta)\,\,,
\end{equation}
\begin{displaymath}
t_{\alpha\,\gamma}\,=\,\int\,d^{3}\,\vec{r}\,\varphi_{\alpha}^{\star}\,
(\vec{r})\,t\,\varphi_{\gamma}\,(\vec{r})\quad {\rm and}\quad
K_{\alpha\,\gamma}\,=\,t_{\alpha\,\gamma}\,-\,\frac{1}{2}\,\sum\limits_{\beta}\,
v_{\alpha\,\beta\,\beta\,\gamma}
\end{displaymath}
in that basis. Whenever possible, we shall use the
matrix notation
\begin{equation}
H\,=\,K\,\cdot\,\rho\,+\,\frac{1}{2}\,\rho\,\cdot\,v\,\cdot\,\rho
\end{equation}
to avoid cumbersome labels and summations. The unphysical
self--interaction term in (3.6), combined with the kinetic energy
$t_{\alpha\,\gamma}$ into $K_{\alpha\,\gamma}$, arises from
anticommuting creation and annihilation operators in the standard form
of $H$ such that the density operator representation (3.5) follows. We
shall come back to this point in section 4.
One then employs the Gaussian trick in its complex version,
\begin{equation}
\exp\left\{-\frac{i}{2}b\cdot{A}\cdot{b}\right\}=
\sqrt{\det{A}^{-1}}\int\left(\prod\limits_{j=1}^{L}\,
\frac{dx_{j}}{\sqrt{2\pi{i}}}\right)\exp\left\{
+\frac{i}{2}x\cdot{A}^{-1}\cdot{x}-ib\cdot{x}\right\}\,\,,
\end{equation}
to the time development operator (3.2), identifying the density operator
$\rho$ with $b$. Relation (3.8) is valid if matrix $A$ is real,
symmetric and invertible.\\[3ex]
The Gaussian trick (3.8) cannot be applied immediately to the two--body
part of $H$ in equation (3.5), since the one--body and two--body parts of
$H$ do not commute. This non--commutativity is treated the standard way
by dividing the time--interval $(t_{f}-t_{i})$ in (3.2) into $M$
slices of equal length $\epsilon=(t_{f}-t_{i})/M$. Then
we can factorize for $\epsilon\to 0$,
\begin{equation}
\exp\,\left\{-\,\frac{i}{\hbar}\,\epsilon\,H\right\}\,=\,
\exp\,\left\{-\,\frac{i}{\hbar}\,\epsilon\,K\,\cdot\,\rho\right\}\,
\exp\left\{-\,\frac{i}{2\,\hbar}\,\epsilon\,\rho\,\cdot\,v\,\cdot\,\rho\right\}\,\,,
\end{equation}
and we may now apply (3.8) to linearize the second factor of (3.9) in
$\rho$:
\begin{equation}
\exp
\left\{-\frac{i}{2\,\hbar}\,\epsilon\,\rho\cdot{v}\cdot\,\rho\right\}=
\sqrt{\det\,(\frac{\epsilon}{\hbar}\,v^{-1})}\int\left(\prod\limits_{\alpha,\,\gamma}
\frac{d\,\sigma_{\alpha\,\gamma}}{\sqrt{2\,\pi\,{i}}}\right)
\exp\left\{\frac{i}{2\,\hbar}\,\epsilon\,\sigma\cdot
{v}^{-1}\cdot\sigma\,-\,\frac{i}{\hbar}\,\epsilon\,
\sigma\cdot\rho\right\}\,\,. 
\end{equation}
For each operator $\rho_{\alpha\,\gamma}$ we have to introduce a real
variable $\sigma_{\alpha\,\gamma}$.
Choosing
\begin{equation}
\tilde{\sigma}\,=\,\sigma\,\cdot\,v^{-1}\,\,,
\end{equation}
we may also write
\begin{equation}
\exp
\left\{-\,\frac{i}{2\,\hbar}\,\epsilon\,\rho\cdot v\cdot\rho\right\}
\,=\,\sqrt{\det\,(\frac{\epsilon}{\hbar}\,v)}\,\int\left(\prod\limits_{\alpha,\,\gamma}\,
\frac{{d\,\tilde{\sigma}_{\alpha\,\gamma}}}{\sqrt{2\,\pi\,i}}\right)
\exp
\left\{\frac{i}{2\,\hbar}\,\epsilon\,\tilde{\sigma}\cdot v\cdot\tilde{\sigma}\,-\,\frac{i}{\hbar}\,\epsilon\,
\tilde{\sigma}\cdot v\cdot\rho\right\}
\end{equation}
as useful alternative to (3.10). While $\tilde{\sigma}$ has the
quality
of a density, $\sigma$ has that of a potential.\\
Matrix $v$ of equation (3.6) is symmetric in particle coordinates
$\vec{r},\,\vec{r}^{\,'}$ and thus in the label pairs
$(\alpha,\,\gamma)$
and $(\beta,\,\delta)$. Hence $v$ fulfills the symmetry requirement on
$A$ in formula (3.8).\\[3ex]
Repeating the above step for each time--interval, labelled by index $k$
in the following, we can write for the time--evolution operator (3.2)
with
(3.10):
\begin{eqnarray}
\lefteqn{\exp\,\left\{-\frac{i}{\hbar}\,(t_{f}-t_{i})\,H\right\}=
\int\left(\prod\limits^{M}_{k=1}\,\sqrt{\det\,(\frac{\epsilon}{\hbar}\,v^{-1})}\,
\prod\limits_{\alpha,\gamma}\,\frac{d\,\sigma_{\alpha\gamma}\,(k)}
{\sqrt{2\,\pi\,i}}\right)}\nonumber\\
& & \times\,\,\exp\,\left\{\frac{i}{2\,\hbar}\,\epsilon
\sum\limits_{k=1}^{M}\sigma(k)\cdot{v^{-1}}\cdot\sigma\,(k)\right\}
\prod\limits_{k=1}^{M}\exp\left\{-\frac{i}{\hbar}\,\epsilon\left(K+\sigma\,(k)\right)
\cdot\rho\right\}\,\,.
\end{eqnarray}
To prepare the next step we note that from (3.10)
\begin{equation}
\left(\det\,(\frac{\epsilon}{\hbar}\,v^{-1})\right)^{-\frac{1}{2}}\,=\,\int\left(\prod\limits_{\alpha,\,\gamma}\,
\frac{d\,\sigma_{\alpha\,\gamma}}{\sqrt{2\,\pi\,i}}\right)\,\exp
\left\{\frac{i}{2\,\hbar}\,\epsilon\,\sigma\,\cdot\,v^{-1}\,\cdot\,\sigma\right\}\,\,.
\end{equation}
We may now take the limit $\epsilon\to 0,\,M \to\infty$ such that
$\epsilon\,M\,=\,(t_{f}\,-\,t_{i})$ remains finite, replacing in (3.13)
\begin{equation}
\epsilon\,\sum\limits_{k=1}^{M}\,\cdots\,\qquad\longrightarrow\qquad
\int\limits_{t_{i}}^{t_{f}}\,d\,t\,\cdots\,\,\,\,\,\,\,\,\,\,\,\,\,.
\end{equation}
Then (3.13) reads as functional integral, using the Trotter formula,
\begin{eqnarray}
\lefteqn{\exp\,\left\{-\,\frac{i}{\hbar}(t_{f}-t_{i})H\right\}\,=
\frac{1}{{\cal{N}}}\,\int\left(\prod\limits_{\alpha,\gamma}\,\frac{D\,\sigma_{\alpha\gamma}}{\sqrt{2\pi\,i}}\right)}\\ & & \times\,\exp\left\{\frac{i}{2\,\hbar}\int\limits_{t_{i}}^{t_{f}}\,dt\,\sigma\,(t)\cdot{v^{-1}}\cdot\sigma\,(t)\right\}
 \,{\cal{T}}\,[\exp\left\{ \,-\frac{i}{\hbar}\int\limits^{t_{f}}_{t_{i}}\, dt\,(K\, +\, \sigma\,(t))\,\cdot\rho\right\}]\nonumber
\end{eqnarray}
where $\cal{T}$ denotes time--ordering and the norm
\begin{equation}
{\cal{N}}\,=\,\int\left(\prod\limits_{\alpha,\,\gamma}\,\frac{D\,\sigma_{\alpha\gamma}}
{\sqrt{2\pi\,i}}\right)\,\exp\,\left\{\frac{i}{2\,\hbar}\,\int\limits_{t_{i}}^{t_{f}}dt
\,\sigma\,(t)\cdot{v^{-1}}\cdot\sigma\,(t)\right\}
\end{equation}
depends still on $T=t_{f}-t_{i}$, i.e.
${\cal{N}}={\cal{N}}(T)$.
As short--hand notation we shall use in the following
\begin{equation}
<\chi^{'}|U\,(T)|\chi>=\frac{1}{\cal{N}}\int{D}\sigma\exp
\left\{\frac{i}{2\,\hbar}\int\,dt\,\sigma\,(t)\cdot{v^{-1}}\cdot\sigma\,(t)\right\}
<\chi^{'}|U_{\sigma}\,(T)|\chi>\,\,,
\end{equation}
\begin{displaymath}
{\cal{N}}\,=\,\int\,D\sigma\,\exp\left\{\frac{i}{2\,\hbar}\int\,dt\,\sigma\,(t)\cdot{v^{-1}}
\cdot\sigma\,(t)\right\}\,\,,
\end{displaymath}
and
\begin{displaymath}
U_{\sigma}(T)\,=\,{\cal{T}}\,[\exp\left\{-\,\frac{i}{\hbar}\,\int\,dt\,h_{\sigma}\,(t)
\right\}]
\end{displaymath}
with single--particle Hamiltonian
\begin{equation}
h\,\left(\sigma\,(t)\right)\,=\,h_{\sigma}\,(t)\,=\,(K\,+\,\sigma\,(t))\,\cdot\,\rho\,\,.
\end{equation}
The time--ordering operator $\cal{T}$ is necessary since in general
\begin{equation}
[h_{\sigma}(t),\,h_{\sigma^{'}}(t^{'})]\,\not=\,0\,\,.
\end{equation}
In (3.18) we have represented the time--evolution operator of a system
with
time--\underline{independent} two--body interaction as superposition of
time--evolution operators with time--\underline{dependent} single--particle
Hamiltonian $h_{\sigma}(t)$ defined through some collective field
$\sigma(t)$. The superposition, being written as functional integral,
contains the typical Gaussian weight factor in the field
$\sigma$.

\section{Ambiguities of the auxiliary field}
There are two sources of ambiguity or, positively put, of freedom in the
choice of the auxiliary field $\sigma\,(t)$ [3, 9]:\\[3ex]
1. There are various ways to divide the Hamiltonian into one-- and
two--body parts,
\setcounter{equation}{0}
\renewcommand{\theequation}{4.\arabic{equation}}
\begin{equation}
H\,=\,T\,+\,V\,=\,(T\,+\,U)\,+\,(V\,-\,U)\,=\,H_{0}\,+\,H^{'}\,\,,
\end{equation}
a fact well--known from the static shell model. All such versions are
equivalent in an exact treatment, however, approximate schemes of
calculating physical quantities will obviously lead to different
results.\\[3ex]
With the choice (3.4) for the density operator $\rho$, the mean field
$\sigma$, determined in lowest order of the stationary phase
approximation to the functional integral (3.18), is of Hartree type.
The Fock
term appears only after quadratic corrections are taken into account. In
contrast, choosing
$\overline{\rho}_{\alpha\,\delta}\,=\,a^{\dagger}_{\alpha}\,a_{\delta}$
as density operator in (3.5), one obtains in lowest order the Fock--term
only which in turn is corrected in second order by the Hartree--term.
With $\Delta^{\dagger}_{\alpha\,\beta}\,=\,a^{\dagger}_{\alpha}\,
a_{\beta}^{\dagger}$ and
$\Delta_{\delta\,\gamma}\,=a_{\gamma}\,a_{\delta}$
it is possible to introduce pairing fields in lowest order.\\[3ex]
All told, the solutions $\sigma$ of the stationary phase approximation
just define a \underline{starting point} for higher orders
of the stationary phase approximation. Their
convergence rate will, of course, depend on the choice of the starting
point.\\[3ex]
2. The second ambiguity lies at the core of the functional integral
representation introduced in section 3. In the expansion of the second
term in the exponent of (3.12),
\begin{equation}
\exp\,\left\{-\,\frac{i}{\hbar}\,\epsilon\,\tilde{\sigma}\,\cdot\,v\,\cdot\,\rho\right\}\,=\,1\,-\,
\frac{i}{\hbar}\,\epsilon\,\tilde{\sigma}\,\cdot\,v\,\cdot\,\rho\,-\,\frac{\epsilon^{2}}{2\,\hbar^{2}}\,(\tilde{\sigma}\,\cdot\,v\,\cdot\,
\rho)^{2}\cdots\,\,\,,
\end{equation}
the linear term does not contribute to the integral as it is odd. The
contribution of (4.2) to the integral in (3.12) stems from the quadratic term
in (4.2) which actually is of order $\epsilon$, since with the Gaussian
weight factor the dominant values of $\tilde{\sigma}$ are of
order
$\epsilon^{-{\frac{1}{2}}}$. We may, therefore, arbitrarily modify the
coefficient of the term linear in $\tilde{\sigma}$ in (4.2),
without changing
the value of the integral (3.12). If we choose this coefficient as
\begin{equation}
W_{\alpha\,\beta\,\gamma\,\delta}\,=\,v_{\alpha\,\beta\,\gamma\,\delta}\,-\,
v_{\alpha\,\beta\,\delta\,\gamma}\,\,,
\end{equation}
then we will have included both Hartree and Fock terms already in lowest
order stationary phase approximation (SPA).\\[3ex]
This linear term is important when one determines the stationary value
$\tilde{\sigma}^{\circ}$ of the integrand in (3.12), but does
not contribute to
the integral (3.12) or finally the functional integral in $\tilde{\sigma}$ corresponding to (3.18). Conversely the
$\tilde{\sigma}$--quadratic terms
in (4.2) do not influence the stationary value of
$\tilde{\sigma}$, however,
they must be taken into account when the functional integral over the
fluctuations around $\tilde{\sigma}^{\circ}$ is calculated.

\section{Stationary phase approximation}
\setcounter{equation}{0}
\renewcommand{\theequation}{5.\arabic{equation}}
To evaluate the integrals in (3.18), we shall use the stationary phase
approximation (SPA). We illustrate the method for a real,
one--dimensional integral
\begin{equation}
I\,(l)\,=\,\int\limits_{-\infty}^{\infty}\,dx\,\exp\,(-f(x)\,/\,l)\,\,.
\end{equation}
For small $l$ the integral is dominated by the stationary points of
$f(x)$. If there is only one solution $f^{'}(x_{0})=0$, then
expansion of
$f(x)$ upto second order in $(x-x_{0})$ gives
\begin{equation}
I\,(l)\,=\,\exp\,(-\,f(x_{0})\,/\,l)\,\left\{\int\limits_{-\infty}^{\infty}\,dx\,\exp\,
\left(-\,\frac{(x\,-\,x_{0})^{2}}{2\,l}\,f^{''}(x_{0})\,\pm\,\cdots\right)\right\}
\end{equation}
\begin{displaymath}
\approx\,\exp\,(-\,f(x_{0})\,/\,l)\,\sqrt{\frac{2\,\pi\,l}{f^{''}(x_{0})}}\,\,,
\end{displaymath}
assuming a minimum $(f^{''}(x_{0})>0)$.
The complex variant of (5.1), (5.2), extended to the multi--dimensional
case, will be applied to the integral (3.18).
The role of the small parameter $l$ is then taken by $\hbar$.
Hence the use of (5.2) implies a \underline{semi--classical}
expansion.\\[3ex]
Rewriting (3.18) as
\begin{equation}
<\chi^{'}|U\,(T)|\chi>\,=\,\frac{1}{\cal{N}}\,\int\,D\sigma\,\exp\,\left\{\frac{i}{\hbar}\,S\,
[\sigma(t)]\right\}
\end{equation}
with
\begin{equation}
S\,=\,\frac{1}{2}\,\int\limits^{t_{f}}_{t_{i}}\,dt\,\sigma\,(t)
\,\cdot\,v^{-1}\,\cdot\,\sigma\,(t)\,-\,i\,\hbar\,\ln <\chi^{'}|{\cal{T}}\,
[\exp (-\,\frac{i}{\hbar}\,\int\limits_{t_{i}}^{t_{f}}\,dt\,h_{\sigma}(t))]|\chi>\,\,,
\end{equation}
the saddle point condition for the functional $S$,
\begin{equation}
\frac{\delta\,S}{\delta\,\sigma_{\alpha\,\gamma}(t)}\,=\,0\quad
{\rm for\,all}\quad \alpha,\,\gamma\,\,,
\end{equation}
reads explicitly, in matrix notation,
\begin{equation}
\sigma^{\circ}\,(t)\,\cdot\,v^{-1}\,=\,
\frac{<\chi^{'}|U_{\sigma^{\circ}}\,(t_{f},\,t)\,\rho\,U_{\sigma^{\circ}}\,(t,\,t_{i})|\chi>}
{<\chi^{'}|U_{\sigma^{\circ}}\,(t_{f},\,t_{i})|\chi>}
\end{equation}
with $U_{\sigma^{\circ}}\,(t^{''},\,t^{'})$ the time--development
operator
under the single--particle Hamiltonian $h_{\sigma^{\circ}}\,(t)$. Note that $\sigma^{\circ}$ will in general be \underline{complex}, although the $\sigma$--fields were originally introduced as \underline{real} integration variables. If we
take
$\chi^{'},\,\chi$ as Slater determinants and generate
$\chi^{'}$ as
\begin{equation}
|\chi^{'}>\,=\,U_{\sigma^{\circ}}\,(t_{f},\,t_{i})|\chi>\,\,,
\end{equation}
then (5.6) simplifies to the TDHF mean field
\begin{equation}
\sigma^{\circ}\,(t)\,=\,\frac
{<\chi|U_{\sigma^{\circ}}\,(t_{i},\,t)\,\rho\,U_{\sigma^{\circ}}\,
(t,\,t_{i})|\chi>}{<\chi|\chi>}\,\cdot\,v
\end{equation}
which refers to the initial time $t_{i}$ only.\\[3ex]
The value of the functional integral (5.3) with action $S$ from (5.4)
is in lowest order
\begin{equation}
<\chi^{'}|U(T)|\chi>\,=\,\exp\left\{\frac{i}{2\,\hbar}\int\,dt\,\sigma^{\circ}\,(t)\cdot{v}^{-1}
\cdot\sigma^{\circ}\,(t)\right\}<\chi^{'}|U_{\sigma^{\circ}}(T)|\chi>\,\,.
\end{equation}
The norm ${\cal{N}}(T)$ is canceled by
the quadratic correction term corresponding to (5.2), 
if we ignore the second term in $S$, eq. (5.4). This term
carries an $i\,\hbar$--factor which makes it weakly oscillating compared
to the first term. This term, dropped in (5.9), generates the Fock term
if one starts from the original version of equations (3.4) and (3.5)
with
Hartree term only, and it cancels the self--interaction term in (3.6). It
also reproduces correlations of the random phase approximation [3].\\[3ex]
Taking $\chi,\,\chi^{'}$ as Slater determinants, the equation of motion
for the density operator is obtained from (3.18), (3.19) and (5.6) by defining Slater
determinants
\begin{equation}
|\Psi\,(t)>\,=\,U_{\sigma^{\circ}}(t,\,t_{i})|\chi>\quad;\quad
<\tilde{\Psi}(t)|\,=\,<\chi^{'}|U_{\sigma^{\circ}}\,(t_{f},\,t)\,\,.
\end{equation}
The time dependence of the respective single--particle functions
$\psi_{i}\,(t),\,\tilde{\psi}_{i}\,(t)$ is then governed by
\begin{equation}
i\,\,\hbar\,\frac{\partial}{\partial\,t}\,|\,\psi_{i}(t)>\,=\,h_{\sigma^{\circ}}(t)|\psi_{i}\,(t)>\quad;\quad
-\,i\,\hbar\,\frac{\partial}{\partial\,t}\,<\,\tilde{\psi}_{i}(t)|\,=\,
<\,\tilde{\psi}_{i}\,(t)|h_{\sigma^{\circ}}\,(t)\,\,
\end{equation}
with time boundary conditions
\begin {equation}
\renewcommand{\theequation}{5.12}
\psi_i\, (t_i)\, = \, \chi_i \,\,;\,\,\tilde{\psi}_{i}(t_f)\,=\,\chi'_i\quad .
\end{equation}
For the density operator
\begin{equation}
\renewcommand{\theequation}{5.13}
\rho\,=\,\sum\limits_{i\,=\,1}^{N}\,\frac{|\,\psi_{i}\,(t)\,><\,\tilde{\psi}_{i}\,(t)\,|}{<\,\tilde{\psi}_{i}\,(t)|\psi_{i}\,(t)\,>}
\end{equation}
one then immediately confirms eq. (1.7),
\begin{equation}
\renewcommand{\theequation}{5.14}
i\,\hbar\,\frac{\partial}{\partial\,t}\,\rho\,=\,[h_{\sigma^{\circ}}\,(t),\,\rho]\,\,.
\end{equation}
Equation (5.14) has to be solved under time--\underline{boundary}
conditions, at $t\,=\,t_{i}$ and $t_{f}$, rather than as
\underline{initial value} problem like (1.1), (1.2).
Each transition $\chi\,\to\,\chi^{'}$ will have its own mean field, and
this allows the method to describe \underline{exclusive} reactions in
contrast to TDHF which at best can be used for \underline{inclusive}
reactions. It is also worth noting that TDMF needs two sets of
single--particle wave functions, $\psi_{i}\,(t)$ and
$\tilde{\psi}_{i}(t)$,
while TDHF is formulated in just one set which develops from the initial
state in the mean field $u$.

\section{Static approximation}
\setcounter{equation}{0}
\renewcommand{\theequation}{6.\arabic{equation}}
On the exact many--body level the connection between the resolvent and
the time--development operator is simply a Fourier transformation
\begin{equation}
<\chi^{'}|(E\,+\,i\,\kappa\,-\,H)^{-1}|\chi>\,=\,-\frac{i}{\hbar}\,\int\limits_{0}^{\infty}\,dT\,\exp\left\{\frac{i}{\hbar}\,(E\,+\,i\,\kappa)\,T\right\}
\,<\chi^{'}|\exp \left(-\frac{i}{\hbar}\,H\,T\right)|\chi>
\end{equation}
for infinitesimal $\kappa>0$ and with $\chi,\,\chi^{'}$ as
initial
and final states. On the mean field level the connection between the resolvent and time--development operator is more intricate due to the non--linear nature of
self--consistent mean field methods.\\[3ex]
The basic assumption to be made is that $\sigma\,(t)$ and hence
$h_{\sigma}\,(t)$ vary slowly in time so that we may approximately
neglect their time dependence. In this static approximation we have from
(3.18) and (3.19)
\begin{equation}
<\chi^{'}|\exp\left\{-\frac{i}{\hbar}\,H\,T\right\}|\chi>=\frac{1}{{\cal{N}}(T)}
\int{D}\sigma\,\exp\left\{\frac{i}{2\hbar}(\sigma\cdot{v^{-1}}\cdot\sigma)\,T\right\}
<\chi^{'}|\exp\left\{-\frac{i}{\hbar}\,h_{\sigma}\,T\right\}|\chi>
\end{equation}
so that (6.1) reads:
\begin{equation}
<\chi^{'}|(E\,+\,i\,\kappa\,-\,H)^{-1}|\chi>\,=\end{equation}\begin{displaymath}-\frac{i}{\hbar}\,\int\limits_{0}^{\infty}dT
\exp\left\{\frac{i}{\hbar}\,(E+i\kappa)T\right\}\frac{1}{{\cal{N}}(T)}\int
D\sigma\exp\left\{\frac{i}{2\hbar}(\sigma\cdot v^{-1}\cdot\sigma)T\right\}
<\chi^{'}|\exp\left\{-\frac{i}{\hbar}\,h_{\sigma}T\right\}|\chi>\,\,.
\end{displaymath}
In $N$--particle Hilbert space with
\begin{equation}
h_{\sigma}\,=\,\sum\limits_{m\,=\,1}^{N}\,h_{m}\,(\sigma)\,\,,
\end{equation}
we can use the inverse Fourier transformation on the single--particle
level,
\begin{equation}
\exp\left\{-\,\frac{i}{\hbar}(h_{m}\,-\,i\,\kappa_{m})\,T\right\}\,=\,\frac{i}{2\pi}\int\limits_{-\infty}^{\infty}\,dw_{m}\,\,
\frac{\exp\,\left\{-\frac{i}{\hbar}\,w_{m}\,T\right\}}{w_{m}\,-\,h_{m}\,+\,i\,\kappa_{m}}
\quad{\rm with}\quad T\,>\,0
\end{equation}
for particle $m$, to obtain an expression for the many--body resolvent in
terms of single--particle resolvents
$g_{m}=(w_{m}-h_{m}+i\kappa_{m})^{-1}$
\begin{eqnarray}
\lefteqn{<\chi^{'}|(E\,+\,i\,\kappa\,-\,H)^{-1}|\chi>\,=
 -\frac{i}{\hbar}\left(\frac{i}{2\pi}\right)^{N}\int\limits_{0}^{\infty}\,dT\frac{1}{{\cal{N}}(T)}\int{D}\sigma\int
\left(\prod\limits_{m=1}^{N}\,dw_{m}\right)}\\ & & \times\,\exp\left\{\frac{i}{\hbar}(E+i\kappa^{'}-
\sum\limits_{m}w_{m}+\frac{1}{2}\sigma\cdot{v^{-1}}\cdot\sigma)T\right\}<\chi^{'}|\prod\limits_{m\,=\,1}^{N}\,(w_{m}\,-\,h_{m}\,+\,i\,\kappa_{m})^{-1}
|\chi>\nonumber
\end{eqnarray}
with the constraint
$\kappa^{'}=\kappa-\sum\limits_{m}\kappa_{m}>0$
to ensure the existence of the $T$--integral.\\[3ex]
The product of single--particle operators in (6.6) acting on the Slater
determinant $\chi$ (to the right or on $\chi^{'}$ to the left) gives
\begin{equation}
\left(\prod\limits_{m\,=\,1}^{N}\,g_{m}\right)\,{\cal{A}}|\chi_{1}\,\chi_{2}\,.\,.\,.\,\chi_{N}>\,=\,{\cal{A}}|
(g_{1}\,\chi_{1})\,(g_{2}\,\chi_{2})\,.\,.\,.\,(g_{N}\,\chi_{N})>
\end{equation}
since the antisymmetrizer ${\cal{A}}$ commutes with the symmetrical
product
$\prod\limits_{m}\,g_{m}$. Equation (6.7) suggests to introduce two sets
of single--particle wave functions
\begin{equation}
\mid\varphi_{m}>\,=\,g_{m}\mid\chi_{m}>\,\,\,;\,\,\,<\varphi^{'}_{m}\mid
\,=\,<\chi^{'}_{m}\mid g_{m}
\end{equation}
or equivalently
\begin{equation}
(w_{m}\,-\,h)\mid\varphi_{m}\,(\sigma,\,w_{m})>\,=\,\mid\chi_{m}>\,\,\,;\,\,\,
<\varphi_{m}^{'}\,(\sigma,\,w_{m})\mid (w_{m}\,-\,h)\,=\,<\chi^{'}_{m}\mid\,\,.
\end{equation}
Note that on the single--particle level we may now drop the label $m$ on
the Hamiltonian $h$
being the same for all (identical!) particles. In this representation
\begin{equation}
<\chi^{'}|(E+i\kappa-H)^{-1}|\chi>=-\frac{i}{\hbar}\left(\frac{i}{2\,\pi}\right)^{N}\int\limits_{0}^{\infty}
dT\int{D}\sigma\int\left(\prod\limits_{m}dw_{m}\right)\frac{1}{{\cal{N}}(T)}\exp\left\{
\frac{i}{\hbar}S\left[\sigma,\,w_{m},\,T\right]\right\}
\end{equation}
with
\begin{equation}
S\,=\,\left(E\,+\,i\,\kappa^{'}\,+\,\frac{1}{2}\,\sigma\,\cdot\,v^{-1}\,\cdot\,\sigma\,-\,
\sum\limits_{m}\,w_{m}\right)\,T\,-\,i\,\hbar\,\sum\limits_{m}\,\ln<\chi^{'}_{m}\mid\varphi_{m}>\,\,.
\end{equation}
In the last term in (6.11) we can replace $<\chi_{m}^{'}\mid\varphi_{m}>$
by
$<\varphi_{m}^{'}\mid\chi_{m}>$ using the bra-- rather than the ket--equation
in (6.8). In both versions we assume, with the same arguments as in
section 2, that the overlap matrices $<\chi_{m}^{'}\mid\varphi_{n}>$ and
$<\varphi_{m}^{'}\mid\chi_{n}>$ are chosen diagonal.
Eqs. (6.5) and (6.8), (6.9) establish a Fourier transformation on the
single--particle level between $\psi_{m}\,(T)$ and
$\varphi_{m}\,(w_{m})$:
\begin{eqnarray}
|\psi_{m}\,(T)> & = & \exp\,\left\{-\,\frac{i}{\hbar}\,(
h\,-\,i\,\kappa_{m})\,T\right\}\,|\,\chi_{m}>\nonumber\\
\nonumber\\
& = & \frac{i}{2\pi}\,\int\limits_{-\,\infty}^{\infty}\,d\,w_{m}\,
\frac{\exp\,\left\{-\,\frac{i}{\hbar}\,w_{m}\,T\right\}}
{w_{m}\,-\,h\,-\,i\,\kappa_{m}}\,|\,\chi_{m}>\nonumber\\
\nonumber\\
& = & \frac{i}{2\,\pi}\,\int\limits_{-\,\infty}^{\infty}\,d\,w_{m}\,
\exp\,\left\{-\,\frac{i}{\hbar}\,w_{m}\,T\right\}\,|\,\varphi_{m}\,(w_{m})>\,\,.
\end{eqnarray}
Equations (6.9) have, apart from the normalization factors $\lambda_i,\,\lambda_i^{'}$, the formal structure of (2.7). However, energies
$w_{m}$ were introduced as \underline{real} variables in contrast to the
\underline{complex} $\eta_{i}$ of (2.7), and $h$ in (6.9) is defined in
terms of the $\sigma$--fields rather than through
$\varphi_{m},\,\varphi_{m}^{'}$. Only after we have solved all
integrations of (6.10) in SPA, we will recover the inhomogeneous TIMF
equations (2.7) together with the correct value of the resolvent matrix
element (2.9).

\section{Path and energy integrals}
These integrals will be solved simultaneously in SPA, followed by the
$T$--integration. While the energies $w_{m}$ and auxiliary fields
$\sigma_{\alpha\,\gamma}$ are introduced as \underline{real,
independent}
variables, their stationary values will be \underline{complex} and
\underline{coupled} through self--consistency.\\[3ex]
The stationarity
equations with respect to $\sigma$ and $w_{m}$ read:
\setcounter{equation}{0}
\renewcommand{\theequation}{7.\arabic{equation}}
\begin{equation}
\frac{\delta\,S}{\delta\,\sigma}\,=\,T\,\sigma\,\cdot\,v^{-1}\,-\,i\,\hbar\,
\sum\limits_{m\,=\,1}^{N}\,\frac{<\chi_{m}^{'}\mid\frac{\delta}{\delta\,\sigma}\mid\varphi_{m}>}
{<\chi_{m}^{'}\mid\varphi_{m}>}\,=\,0
\end{equation}
\begin{equation}
\frac{\partial\,S}{\partial\,w_{n}}\,=\,-T\,\,-i\,\hbar\,\sum\limits^{N}_{m\,=\,1}\,
\frac{<\chi^{'}_{m}\mid\frac{\partial}{\partial\,w_{n}}\mid\varphi_{m}>}
{<\chi_{m}^{'}\mid\varphi_{m}>}\,=\,0\,\,.
\end{equation}
The above derivatives of $\varphi_{m}$ can be calculated with the help
of equations (6.9): Multiplying
\begin{equation}
(w_{m}\,-\,h)\,\frac{\delta}{\delta\,\sigma}\,\mid\varphi_{m}>\,-\,\frac
{\delta\,h}{\delta\,\sigma}\mid\varphi_{m}>\,=\,0
\end{equation}
by $<\varphi_{m}^{'}\mid$ gives, with $h$ from (3.19),
\begin{equation}
<\chi^{'}_{m}\mid\frac{\delta}{\delta\,\sigma}\mid\varphi_{m}>\,=\,
<\varphi_{m}^{'}\mid\rho\mid\varphi_{m}>\,\,.
\end{equation}
Explicitly:
\begin{equation}
<\chi^{'}_{m}\mid\frac{\delta}{\delta\,\sigma_{\alpha\,\gamma}}\mid\varphi_{m}>\,=\,
<\varphi_{m}^{'}\mid a^{\dagger}_{\alpha}\,a_{\gamma}\mid\varphi_{m}>\,=\,
<\varphi_{m}^{'}\mid\alpha><\gamma\mid\varphi_{m}>\,\,.
\end{equation}
Derivatives of (6.9) with respect to $w_{n}$ give
\begin{equation}
\delta_{nm}\,\left\{\mid\varphi_{m}>\,+\,(w_{m}\,-\,h)\,\frac{\partial}
{\partial\,w_{n}}\mid\varphi_{m}>\right\}\,=\,0\,\,.
\end{equation}
Multiplying by $<\varphi_{m}^{'}\mid$ and using again (6.9) leads to
\begin{equation}
<\chi^{'}_{m}\mid\frac{\partial}{\partial\,w_{n}}\mid\varphi_{m}>\,=\,-\,
\delta_{m\,n}<\varphi_{m}^{'}\mid\varphi_{m}>\,\,.
\end{equation}
We insert relations (7.4) and (7.7) into (7.1) and (7.2) to find the
explicit stationarity equations. The stationary value of $\sigma$ is
given by
\begin{equation}
\sigma^{\circ}\,\cdot\,v^{-1}\,=\,\frac{i\,\hbar}{T}\,\sum\limits_{m}\,
\frac{<\varphi_{m}^{'}\mid\rho\mid\varphi_{m}>}{<\chi_{m}^{'}\mid\varphi_{m}>}
\end{equation}
or
\begin{equation}
\sigma^{\circ}\,=\,\frac{i\,\hbar}{T}\,\sum\limits_{m}\,
\frac{<\varphi_{m}^{'}\mid\rho\mid\varphi_{m}>\,\cdot\,v}
{<\chi_{m}^{'}\mid\varphi_{m}>}\,\,.
\end{equation}
Explicitly with labels of single--particle states
\begin{equation}
\sigma_{\alpha\,\gamma}^{\circ}\,=\,\frac{i\,\hbar}{T}\,\sum\limits_{m\atop
\beta,\,\delta}
\,\frac{<\varphi_{m}^{'}\mid\beta><\delta\mid\varphi_{m}>}
{<\chi_{m}^{'}\mid\varphi_{m}>}\,v_{\beta\,\alpha\,\delta\,\gamma}\,
=\,\frac{i\,\hbar}{T}\,\sum\limits_{m}\,
\frac{<\varphi_{m}^{'}\,\alpha\mid v\mid\varphi_{m}\,\gamma>}
{<\chi_{m}^{'}\mid\varphi_{m}>}
\end{equation}
using the completeness of single--particle states $\beta,\,\delta$
referring to the representation (3.5) of the Hamiltonian $H$. The
stationary value
$w_{m}^{\circ}$ is determined implicitly through
\begin{equation}
\frac{<\varphi_{m}^{'}\mid\varphi_{m}>}{<\chi_{m}^{'}\mid\varphi_{m}>}
\,=\,-\,\frac{i}{\hbar}\,T
\end{equation}
or, using (6.9) again, explicitly by
\begin{equation}
\frac{i\,\hbar}{T}\,=\,\frac{<\chi_{m}^{'}\mid\varphi_{m}>}
{<\varphi_{m}^{'}\mid\varphi_{m}>}\,=\,
\frac{<\varphi_{m}^{'}\mid(w_{m}^{\circ}\,-\,h)\mid\varphi_{m}>}
{<\varphi_{m}^{'}\mid\varphi_{m}>}\,=\,w^{\circ}_{m}\,-\,\varepsilon_{m}^{\circ}
\end{equation}
with
\begin{equation}
\varepsilon_{m}^{\circ}\,=\,\frac{<\varphi_{m}^{'}\mid h\,(\sigma^{\circ})\mid\varphi_{m}>}{
<\varphi_{m}^{'}\mid\varphi_{m}>}\,\,.
\end{equation}
After $T$--integration in section 8, $E$ rather than $T^{-1}$ will
appear
in relation (7.12). Combining (7.10) and (7.11) identifies
$\sigma^{\circ}$ as \underline{Hartree--Fock} field,
\begin{equation}
\sigma_{\alpha\,\gamma}^{\circ}\,=\,\sum\limits_{m\,=\,1}^{N}\,
\frac{<\varphi_{m}^{'}\,\alpha\mid v\mid\varphi_{m}\,\gamma>}
{<\varphi_{m}^{'}\mid\varphi_{m}>}\,\,\,,
\end{equation}
since according to section 4 the matrix $v$ is assumed to be
antisymmetrized.
A slight modification of (4.3) also allows to eliminate the unphysical
self interaction in (3.6) so that $K$ reduces to the kinetic energy
operator.\\[3ex]
The normalization factors in (6.10), ${\cal{N}}(T)$ and
$(i\,/\,2\,\pi)^{N}$, are eliminated by the Gauss correction to the
lowest order SPA which approximates the integrals by the integrand
taken at the stationary point $(\sigma^{\circ},\,w^{\circ})$. The
Gauss correction to the $\sigma$--integral,
\begin{equation}
\left(
\det\,\left(\frac{\delta^{2}\,S}{\delta\sigma\,\delta\sigma^{'}}\right)_{\sigma^{\circ}}
\right)^{-\frac{1}{2}}\,=\,\left(
\det\,(T\,v^{-1})\right)^{-\frac{1}{2}}\,\,\,,
\end{equation}
cancels ${\cal{N}}(T)$ according to (3.14), if the $\sigma$--dependence
of the
ln--term in (6.11) is neglected.
This ln--term carries a factor
$\hbar$ as compared to the leading term. Hence the above assumption is
justified in semi--classical approximation. For the Gauss correction to
the $w$--integration, each integral can be treated separately as the
integrand factorizes. From (7.2) and (7.7), together with (6.9) and
(7.12),
it follows that the leading term in the second derivative of $S$ with
respect to $w_{m}$ is given by
\begin{equation}
\left(\frac{\partial^{2}\,S}{\partial\,w_{m}^{2}}\right)_{w^{\circ}_m}\,=\,
\frac{i}{\hbar}\,T^{2}\,\,\,,
\end{equation}
and the total correction factor is
\begin{equation}
\sqrt{\prod\limits_{m}\,\left(
\frac{2\,\pi\,i\,\hbar}{S^{''}(w_{m})}\right)_{w^{\circ}_m}}\,=\,
\left(\sqrt{2\,\pi}\,\,\frac{\hbar}{T}\right)^{N}\,\,.
\end{equation}
The correction to (7.16) involves fluctuations of the single--particle
propagators which go beyond the mean field concept.\\[3ex] To this order
the matrix element
(6.10) of the resolvent reads
\begin{equation}
D\,(E\,+\,i\,\kappa)\,=\,<\chi^{'}|(E\,+\,i\,\kappa\,-\,H)^{-1}|\chi>\,=
\end{equation}
\begin{displaymath}
-\frac{i}{\hbar}\left(\frac{e}{{\sqrt{2\pi}}}\right)^{N}\int\limits_{0}^{\infty}dT
\left(\frac{i\hbar}{T}\right)^{N}\exp\left\{\frac{i}{\hbar}
\left(E+\frac{1}{2}\,\sigma^{\circ}\cdot{v^{-1}}\cdot\sigma^{\circ}-
\sum\limits_{m}\varepsilon_{m}^{\circ}+i\kappa^{'}\right)T\right\}
<\chi^{'}\mid\phi>\,\,,
\end{displaymath}
after replacing $w_{m}^{\circ}$ by $\varepsilon_{m}^{\circ}$ according
to (7.12). With $e\,/\sqrt{2\,\pi}=1.084...$, the normalization
factor
$(i\,/\,2\,\pi)$ is not exactly cancelled by the Gaussian correction.
However, this is just a \underline{constant}, \underline{real} factor
which does not harm the analytical properties of the resolvent.

\section{$T$--integration}
With regard to SPA, we extend the range of integration for $T$ to
$-\infty$, introducing the step function in its integral representation,
\setcounter{equation}{0}
\renewcommand{\theequation}{8.\arabic{equation}}
\begin{equation}
\Theta\,(T)\,=\,\frac{1}{2\,\pi\,i}\,\int\limits_{-\infty}^{\infty}\,ds
\,\frac{e^{\frac{i}{\hbar}\,s\,T}}{s\,-\,i\,
\kappa^{''}} \quad {\rm for} \,\,\, \kappa^{''}\,>\,0\,\,.
\end{equation}
Then, with $e\,/\sqrt{2\,\pi}\approx 1$, we have
\begin{equation}
D\,(E\,+\,i\,\kappa)\,=\,\frac{1}{2\,\pi\,\hbar}\,\int\limits_{-\infty}^{\infty}\,dT\,
\int\limits_{-\infty}^{\infty}\,ds\,\exp\left\{\frac{i}{\hbar}\,S[s,\,T]\right\}
\end{equation}
with
\begin{eqnarray}
\lefteqn{S\,[s,\,T]\,=\,\left(E\,+\,\frac{1}{2}\,\sigma^{\circ}\,\cdot\,v^{-1}\,\cdot\,
\sigma^{\circ}
\,-\,\sum\limits_{m}\,\varepsilon_{m}^{\circ}\,+\,s\,+\,i\,\kappa^{'}\right)\,T}\nonumber\\
& & -\,i\,\hbar\,\sum\limits_{m}\,\ln <\chi^{'}_{m}\mid\varphi_{m}>\,-\,i\,\hbar\,N\,\ln\left(
\frac{i\,\hbar}{T}\right)\,+\,i\,\hbar\,\ln\,(-s\,+\,i\,\kappa^{''})\,\,.
\end{eqnarray}
Simultaneous SPA to both integrals gives as stationarity equations
\begin{equation}
\frac{\partial\,S}{\partial\,s}\,=\,T\,+\,
\frac{i\,\hbar}{s\,-\,i\,\kappa^{''}}\,=\,0\quad{\rm or}\quad
-s_{\circ}\,=\,\frac{i\,\hbar}{T_{\circ}}
\end{equation}
and
\begin{equation}
\frac{\partial\,S}{\partial\,T}\,=\,
E\,+\,\frac{1}{2}\,\sigma^{\circ}\,\cdot\,{v^{-1}}\,\cdot\,\sigma^{\circ}\,-\,\sum\limits_{m}
\varepsilon_{m}^{\circ}\,+\,s\,+\,i\,\kappa^{'}\,=\,0\end{equation}\begin{displaymath}\quad{\rm
or}\quad -\,s_{\circ}\,=\,
E\,+\,\frac{1}{2}\,\sigma^{\circ}\,\cdot\,{v^{-1}}\,\cdot\,\sigma^{\circ}\,-\,\sum\limits_{m}
\varepsilon_{m}^{\circ}
\end{displaymath}
where $\kappa^{'},\,\kappa^{''}$ are now superfluous. Eq. (8.5) is
\underline{exact}
though it looks as if in (8.3) only the term proportional to $T$ had been
taken into account. The contributions from the remaining terms cancel
exactly by virtue of eqs. (6.9).\\[3ex]
For the Gauss correction to the lowest order SPA of the $T$-- and
$s$--integrals, one has to calculate the Hesse--matrix $M$ of $S$ with
respect to $T$ and $s$ since $S$ is not additive in the variables $T$
and $s$. One finds
\begin{equation}
M\,=\,\left(
\begin{array}{cc}
0 & 1\\
1 & \frac{i}{\hbar}\,T_{\circ}^{2}
\end{array}
\right)
\end{equation}
with
\begin{equation}
\frac{\partial^{2}\,S}{\partial\,T^{2}}\,=\,0
\end{equation}
in semi--classical approximation. Hence $\det M=-1$, and the Gauss
correction factor for the $T$-- and $s$--integrations is
$\sqrt{-(2\pi\hbar i)^{2}}=2\pi\hbar$, which cancels the
corresponding factor in (8.2). To this order of SPA we have
\begin{eqnarray}
D\,(E\,+\,i\,\kappa) & = &
\exp\,\left\{\frac{i}{\hbar}\,S\,[T_{\circ},\,s_{\circ}]\right\}\,=\,\left(
\frac{i\,\hbar}{T_{\circ}}\right)^{N-1}<\chi^{'}|\phi>\\
& = &
\left(E\,+\,\frac{1}{2}\,\sigma^{\circ}\,\cdot\,v^{-1}\,\cdot\,\sigma^{\circ}\,-\,
\sum\limits_{m}\,\varepsilon^{\circ}_{m}\right)^{N-1}<\chi^{'}|\phi>\,\,,\nonumber
\end{eqnarray}
with $\phi=\phi (\sigma^{\circ},w^{\circ}_{m})$. The same result
is obtained following an alternative line of reasoning: In the above
semi--classical approximation, with $S$ linear in $T$ and derivative
$\,\partial\, S\,/\,\partial\, T\,=\,
E+\newline\frac{1}{2}\sigma^{\circ}\cdot
 v^{-1}\cdot\sigma^{\circ}-\sum\limits_{m}\varepsilon_{m}^{\circ}+s$
independent of $T$, the $T$--integral reduces to a $\delta$--distribution,
\begin{equation}
\frac{1}{2\pi\hbar}\int\limits_{-\infty}^{\infty}dT\exp\left\{\frac{i}{\hbar}\,\Big(
E+\frac{1}{2}\sigma^{\circ}\cdot{v^{-1}}\,\cdot\sigma^{\circ}-\sum\limits_{m}
\varepsilon_{m}^{\circ}+s\Big )T\right\}=\delta \left(
E+\frac{1}{2}\sigma^{\circ}\cdot{v^{-1}}\cdot\sigma^{\circ}-\sum\limits_{m}
\varepsilon^{\circ}_{m}+s \right).
\end{equation}
With $(i\hbar/T)$ from the stationarity condition (8.4),
the remaining $s$--integral can then be solved exactly,
\begin{equation}
\int\limits_{-\infty}^{\infty}d\,s\,\frac
{\delta\,\left(E\,+\,\frac{1}{2}\,\sigma^{\circ}\,\cdot\,v^{-1}\,\cdot\,
\sigma^{\circ}\,-\,\sum\limits_{m}\,\varepsilon_{m}^{\circ}\,+\,s\right)}
{-\,s}\,(-\,s)^{N}<\chi^{'}|\phi>\,=\,(-\,s_{\circ})^{N-1}<\chi^{'}|\phi>
\end{equation}
with $s_{\circ}$ from (8.5), in complete agreement with (8.8).

\section{Conclusions}
With $\sigma^{\circ}$ from stationarity condition (7.14), the
single--particle
Hamiltonian $h\,(\sigma^{\circ})$, eq. (3.19), has precisely the structure of
$h$ of section 2 with expectation value
\setcounter{equation}{0}
\renewcommand{\theequation}{9.\arabic{equation}}
\begin{equation}
\varepsilon^{\circ}_{m}\,=\,
\frac{<\varphi_{m}^{'}\mid(t\,+\,\sigma^{\circ})\mid\varphi_{m}>}{<\varphi_{m}^{'}\mid\varphi_{m}>}\,=\,
\frac{<\varphi_{m}^{'}\mid (t\,+\,u)\mid\varphi_{m}>}{<\varphi^{'}_{m}\mid\varphi_{m}>}\,\,.
\end{equation}
From (7.8), (7.11) and (7.14) one finds
\begin{equation}
\frac{1}{2}\,\sigma^{\circ}\,\cdot\,v^{-1}\,\cdot\,\sigma^{\circ}\,=\,
\frac{1}{2}\,\sum\limits_{m}\,\frac{<\varphi_{m}^{'}\mid\sigma^{\circ}\mid\varphi_{m}>}
{<\varphi_{m}^{'}\mid\varphi_{m}>}\,=\,\frac{1}{2}\,
\sum\limits_{m,\,n}\,\frac{<\varphi_{m}^{'}\,\varphi_{n}^{'}\mid v\mid\varphi_{m}\,\varphi_{n}>}
{<\varphi_{m}^{'}\mid\varphi_{m}><\varphi_{n}^{'}\mid\varphi_{n}>}
\end{equation}
so that with (7.12), (8.4) and (8.5)
\begin{equation}
E\,+\,\frac{1}{2}\,\sigma^{\circ}\cdot{v^{-1}}\cdot\sigma^{\circ}-\sum\limits_{m}
\varepsilon_{m}^{\circ}\,=\,E\,-\,\frac{<\phi^{'}\mid H\mid\phi>}{<\phi^{'}\mid\phi>}\,=\,
{\overline{\eta}}\,=\,w_{m}^{\circ}\,-\,\varepsilon_{m}^{\circ}\,=\,\frac{i\,\hbar}{T_{\circ}}\,\,,
\end{equation}
and we can identify the (complex) stationary value $w_{m}^{\circ}$ with
$\eta_{m}$ of eq. (2.8). Hence eqs. (6.9) and (2.7) differ only by
factors
$\lambda_{i},\,\lambda_{i}^{'}$ and we have to identify
\begin{equation}
\varphi_{m}\,(\sigma^{\circ},\,w_{m}^{\circ})\,=\,\frac{\varphi_{m}}{\lambda_{m}}\,\,\,\,;\,\,\,\,
\varphi_{m}^{'}\,(\sigma^{\circ},\,w_{m}^{\circ})\,=\,\frac{\varphi_{m}^{'}}{\lambda_{m}^{'}}\,\,.
\end{equation}
We make now use of (2.13) to obtain the final result for (8.8)
\begin{equation}
D\,(E\,+\,i\,\kappa)\,=\,<\chi^{'}|\phi>\,=\,<\phi^{'}|\chi>
\end{equation}
which fully agrees with equation (2.9) for the stationary value of
functional $F$ in the TIMF--method.
In summary we can state that we have derived the inhomogeneous
single--particle equations \underline{and} the stationary value of the
Green function
of the TIMF--method from the TDMF in static, semi--classical
approximation, with the Gauss--correction to the lowest order SPA taken into account.\\[3ex]
Under various conceptual and practical criteria, TIMF can be placed
between TDMF and TDHF (see also table 1): Both TDMF and TIMF methods calculate a mean field
for each transition $\chi\to\chi^{'}$. Hence the respective transition
amplitudes and resolvent matrix elements refer to exclusive reactions,
while TDHF is applicable to inclusive reactions only. TDMF is
conceptually superior to TIMF, the latter being a static approximation
of the former method. Alternately put, TIMF is ''local'' in energy whereas
TDMF is ''non--local'' in time. From the
practical point of view, TIMF requires solving inhomogeneous, complex
equations of Hartree--Fock type for given energy. In this respect it is
comparable to TDHF which is ''local'' in time. The problem of TDMF lies in
combining self--consistency with given boundary conditions in time. No
practicable algorithm for this highly ''non--local'' problem seems to exist
for use in actual numerical calculations. The practical advantage of
TIMF becomes even more pronounced when considering the S--matrix. TDMF
then has to calculate three auxiliary fields rather than one [2, 3]
while the corresponding T--matrix of TIMF [6, 7] is obtained again
from inhomogeneous, complex equations of Hartree--Fock type as for the
above resolvent, with only slight generalization of the
inhomogeneities.\\[6ex]
{\bf{Acknowledgements:}}\\
One of us (A.W.) wants to thank the Japan Society for Promotion of Science and the Deutsche Akademische Austauschdienst for financial support during his stay in Japan. He is also very grateful for the hospitality found at the Yukawa Institute for Theoretical Physics in Kyoto and for helpful discussions with Y. Abe and H. Horiuchi.
\\[6ex]
{\bf{References:}}
\begin{enumerate}
\item K. Goeke, R. Y. Cusson, F. Gr\"ummer, P.-G. Reinhard and H.
Reinhardt: Prog. Theor. Phys. [Suppl.] {\bf 74 \& 75}, 33 (1983)
\item H. Reinhardt: Nuclear Physics A {\bf 390}, 70 (1982)
\item J. W. Negele and H. Orland: Quantum Many--Particle Systems
(Addison--Wesley, 1988)
\item H. Reinhardt: Fortschr. d. Physik {\bf 30}, 127 (1982)
\item J. P. Blaizot and G. Ripka: Quantum Theory of Finite Systems (MIT Press, 1986)
\item B. Giraud, M. A. Nagarajan and I. J. Thompson, Ann. Phys. (N. Y.) {\bf 152}, 475 (1984);\\ B. Giraud, M. A. Nagarajan, Ann. Phys. (N. Y.) {\bf 212}, 260 (1991)
\item J. C. Lemm, Ann. Phys. (N. Y.) {\bf 244}, 136 (1995) 
\item A. Wierling, B. Giraud, F. Mekideche, H. Horiuchi, T. Maruyama, A.
Ohnishi, J. C. Lemm and A. Weiguny, Z. Phy. A {\bf 348}, 153 (1994)
\item A. K. Kerman, S. Levit and T. Troudet: Ann. Phys. (N. Y.) {\bf 148}, 436 (1983)
\end{enumerate}
\pagebreak
\thispagestyle{empty}
\begin{center}
{\LARGE Mean Field Methods for Reactions}
\end{center}\vspace{0.7cm}
\makebox[2cm]{}\makebox[9cm]{\Large Time--Dependent}\makebox[6cm]{\Large Time--Independent}\\[3ex]
{\large Basic quantity}\\
\makebox[2cm]{}\makebox[7.5cm][r]{$\,\,\,<\chi'\mid \exp\{-\frac{i}{\hbar}(t_f-t_i)H\}\mid\chi>$}\makebox[3cm]{$\begin{array}{c}{\scriptstyle{\rm Fourier}}\\\Longleftrightarrow \\{\scriptstyle{\rm Transformation}}\end{array}$}\makebox[4.5cm][l]{$<\chi'\mid(E-H)^{-1}\mid\chi>\,\,\,$}\\[2ex]
{\large Mean Field Approximation}\hspace{1.7cm}$\bigg\Downarrow$\hspace{6.2cm}$\bigg\Downarrow$\\
\makebox[1cm]{}\makebox[3cm]{TDHF}\makebox[0.1cm]{$\qquad\qquad\begin{array}{c}{\scriptstyle{{\rm restriction\,\,of\,\, \chi'}}}\\\Longleftarrow\\{\scriptstyle{\mid\chi'>=U_h(t_f-t_i)\mid\chi>}}\end{array}$}\makebox[7cm]{TDMF}\makebox[0.1cm]{$\begin{array}{c}{\scriptstyle{\rm static}}\\\Longrightarrow\\{\scriptstyle{\rm approximation}}\end{array}$}\makebox[6cm]{TIMF}\\ [2ex]
{\large Equation of motion}\\[2ex]
\makebox[1cm]{}\makebox[3cm]{$[h,\rho ]=i\hbar\dot{\rho}$}\makebox[7cm]{$[h,\rho]=i\hbar\dot{\rho}$}\makebox[6cm]{$[h,\rho]=\overline{\eta}(\rho_{\chi'}-\rho_{\chi})$}\\
\makebox[4cm]{}\makebox[7cm]{to be solved for given}\\
\makebox[1cm]{}\makebox[3cm]{$\rho(t_i)$}\makebox[7cm]{$\rho(t_i)\,,\,\rho(t_f)$}\makebox[6cm]{$\chi\,,\,\chi'$}\\[3ex] 
{\large Type of problem}\\[2ex]
\makebox[1cm]{}\makebox[3cm]{initial}\makebox[7cm]{boundary}\makebox[6cm]{inhomogeneous problem}\\
\makebox[0.8cm]{}\makebox[9cm]{condition problem in $t$:}\makebox[1.2cm]{}\makebox[6cm]{parametric in $E$:}\\
\makebox[1cm]{}\makebox[3cm]{''local''}\makebox[7cm]{''non--local''}\makebox[6cm]{''local''}\\[3ex]
{\large Representation of density $\rho$ in basis}\\[2ex]
\makebox[1cm]{}\makebox[3cm]{$\mid\psi_m(t)>$}\makebox[7cm]{$\mid\psi_m(t)>\,,\,<\tilde{\psi}_m(t)\mid$}\makebox[6cm]{$\mid\varphi_m>\,,\,<\varphi_m'\mid$}\\
\makebox[1cm]{}\makebox[3cm]{orthogonal}\makebox[7cm]{biorthogonal}\makebox[6cm]{biorthogonal}\\[3ex]
{\large Single--particle Hamiltonian $h$}\\[2ex]
\makebox[1cm]{}\makebox[3cm]{Hermitian}\makebox[7cm]{non--Hermitian}\makebox[6cm]{non--Hermitian}\\[3ex]
{\large Specification of asymptotic channels}\\[2ex]
\makebox[1cm]{}\makebox[3cm]{no}\makebox[7cm]{yes}\makebox[6cm]{yes}\\[3ex]
{\large Reactions described}\\[2ex]
\makebox[1cm]{}\makebox[3cm]{inclusive}\makebox[7cm]{exclusive}\makebox[6cm]{exclusive}\\[2ex]
\begin{center}
Table 1
\end{center}

\end{document}